\documentclass[twocolumn,aps,prl,superscriptaddress,floatfix,longbibliography,10pt]{revtex4-1}

\usepackage[T1]{fontenc}
\usepackage[latin9]{inputenc}
\usepackage{amsmath}
\usepackage{amssymb}
\usepackage{graphicx}
\usepackage{upgreek}
\usepackage{dsfont}
\usepackage{xcolor}
\usepackage[colorlinks=true,citecolor=blue,linkcolor=blue]{hyperref}

\newcommand{\mathcomma}{\,,}
\newcommand{\mathperiod}{\,.}
\newcommand{\dd}{\mathrm{d}}
\DeclareMathOperator{\tr}{tr}
\DeclareMathOperator{\arsinh}{arsinh}
\newcommand{\angletheta}{\varphi}
\newcommand{\dosf}{\rho_\mathrm{F}}
\newcommand{\vf}{v_\mathrm{F}}
\newcommand{\kf}{k_\mathrm{F}}

\begin{document}

\title{Short-distance breakdown of the Higgs mechanism and the robustness
of the BCS~theory for charged superconductors}

\author{Sonja Fischer }

\affiliation{Institut f\"ur Theorie der Kondensierten Materie, Karlsruher Institut f\"ur Technologie, D-76131 Karlsruhe, Germany}

\affiliation{Institute for Theoretical Physics and Center for Extreme Matter and Emergent Phenomena, Utrecht University, Princetonplein 5, 3584 CC Utrecht, The Netherlands}

\author{Matthias Hecker}

\affiliation{Institut f\"ur Theorie der Kondensierten Materie, Karlsruher Institut f\"ur Technologie, D-76131 Karlsruhe, Germany}

\affiliation{Institut f\"ur Festk\"orperphysik, Karlsruher Institut f\"ur Technologie, D-76021 Karlsruhe, Germany}

\author{Mareike Hoyer }

\affiliation{Institut f\"ur Theorie der Kondensierten Materie, Karlsruher Institut f\"ur Technologie, D-76131 Karlsruhe, Germany}

\affiliation{Institut f\"ur Festk\"orperphysik, Karlsruher Institut f\"ur Technologie, D-76021 Karlsruhe, Germany}

\author{J\"org Schmalian}

\affiliation{Institut f\"ur Theorie der Kondensierten Materie, Karlsruher Institut f\"ur Technologie, D-76131 Karlsruhe, Germany}

\affiliation{Institut f\"ur Festk\"orperphysik, Karlsruher Institut f\"ur Technologie, D-76021 Karlsruhe, Germany}

\date{\today}

\begin{abstract}
Through the Higgs mechanism, the long-range Coulomb interaction eliminates the low-energy Goldstone phase mode in superconductors and
transfers spectral weight all the way up to the plasma frequency.
Here we show that the Higgs mechanism breaks down for length scales
shorter than the superconducting coherence length while it stays intact, even at high energies, in the long-wavelength limit. This effect is
a consequence of the composite nature of the Higgs field of superconductivity and the broken Lorentz invariance in a solid.
Most importantly, the breakdown of the Higgs mechanism inside the
superconducting coherence volume is crucial to ensure the stability of 
the BCS mean-field theory in the weak-coupling limit.
We also show that changes in the gap equation due to plasmon-induced fluctuations can lead to significant corrections to the mean-field theory and reveal that changes in the density-fluctuation spectrum of a superconductor are not limited to the vicinity of the gap. 
\end{abstract}

\maketitle

\section{Introduction}
The Higgs mechanism is one of the corner stones of modern physics~\cite{Englert64,Higgs64,Guralnik64}:
A Higgs field mixes with gauge bosons rendering them massive, while
Goldstone modes disappear. The remaining degree of freedom becomes
the Higgs boson, the fluctuation in the amplitude of the Higgs field,
which is a new scalar particle. This mechanism was first discussed
by Anderson~\cite{Anderson63} and motivated by the theory of collective
excitations in superconductors~\cite{PWA581,PWA582}. Here, massive
transverse photons emerge and give rise to the Meissner effect. At
the same time, low-energy phase fluctuations of the neutral superfluid
disappear as Goldstone modes in the case of charged superconductors.

In the condensed-matter version of the Higgs mechanism, the Higgs field
describes a Cooper pair; that is, it is known to be a composite bound
state and not a fundamental particle. This has profound implications.
The Higgs boson is no sharply defined resonance but a diffusive object
with a branch-cut singularity. It cannot be analyzed without allowing
for a cloud of quasiparticles to be excited simultaneously. This can
be seen explicitly in a clean superconductor with gap $\Delta_{0}$.
Then, $2\Delta_{0}$ is the excitation energy of amplitude fluctuations,
also known as the Higgs mass~\cite{VolkovKogan-JETP1974,LittlewoodVarma-PRL1981,LittlewoodVarma-PRB1982,Varma-JLowTempPhys2002,PekkerVarma-AnnuRevCondMatPhys2015}. However, excitations with energy $\omega>2\Delta_{0}$
will break Cooper pairs and excite individual quasiparticles. 
Thus, a long-distance continuum theory of the Higgs boson is not free of
intricacies. For an interesting discussion of the Higgs mode in disordered superconductors, see Ref.~\onlinecite{CeaEtAl-PRB2014}. 

Within a Lorentz-invariant treatment of the problem, phase fluctuations
of the superconductor are completely eliminated from the theory. In 
models without Lorentz invariance, 
the Goldstone mode is still eliminated and spectral weight is transferred
to the plasma frequency; this situation is more realistic to superconductors. In either case, the spectrum of phase fluctuations
is dramatically altered by the presence of the long-range Coulomb interaction.

\begin{figure*}[t]
\includegraphics[width=\textwidth]{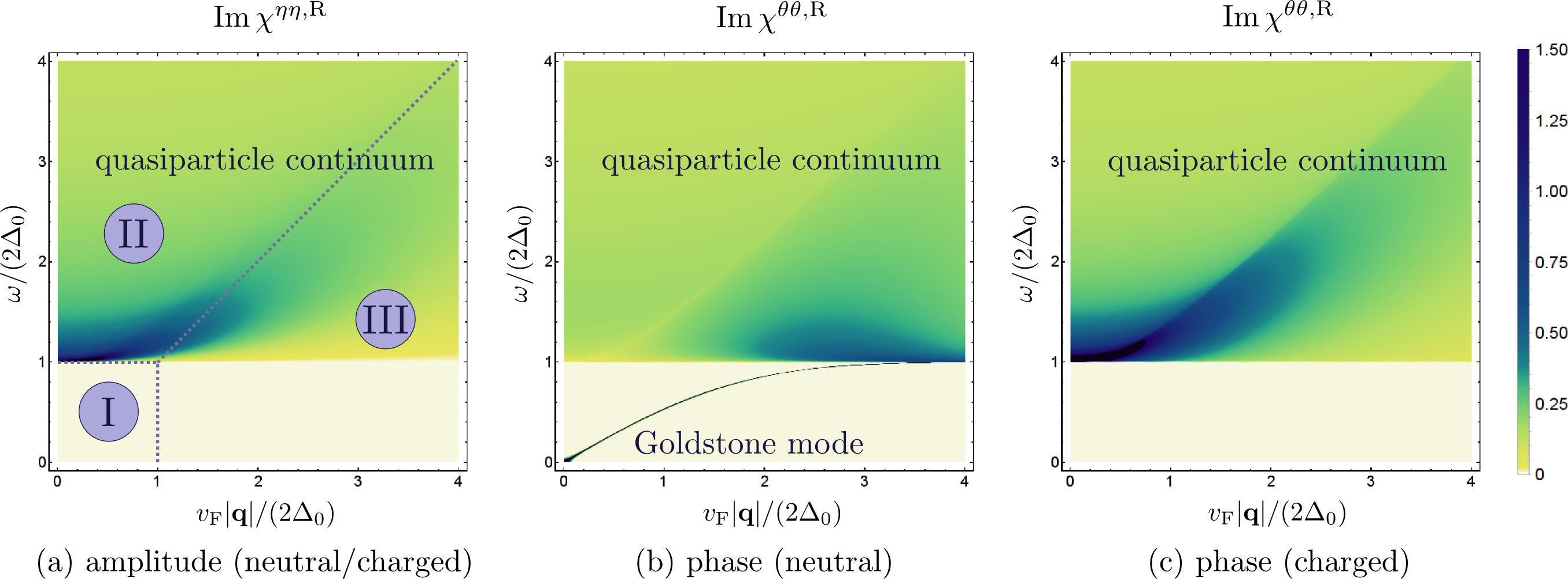} 
\caption{Collective modes of superconductors in $d=3$. (a) shows the amplitude mode
of a superconductor. For particle-hole symmetric systems, i.\,e., with
constant density of states near the Fermi level, the result is the
same for neutral superfluids and charged systems. (b) shows the spectrum
of phase fluctuations of a neutral superfluid with Goldstone mode at low energies $\omega<2\Delta_0$.
(c) shows the phase fluctuation spectrum of a charged superconductor.
While the Goldstone mode disappears through the Higgs mechanism, there
are still collective charge excitations at higher frequencies $\omega>2\Delta_0$.
The coupling to collective charge modes enhances phase fluctuations
near $\omega=2\Delta_0$ compared to the neutral system. A large part
of the spectral weight of phase fluctuations is transferred to the
plasma frequency (not shown). Additionally, in (a), we illustrate the three regimes which are analyzed individually in Secs.~\ref{sec:long-wavelength} and~\ref{sec:short-wavelength}.}
\label{fig:density_plots}
\end{figure*}

\begin{figure*}
\includegraphics[width=\textwidth]{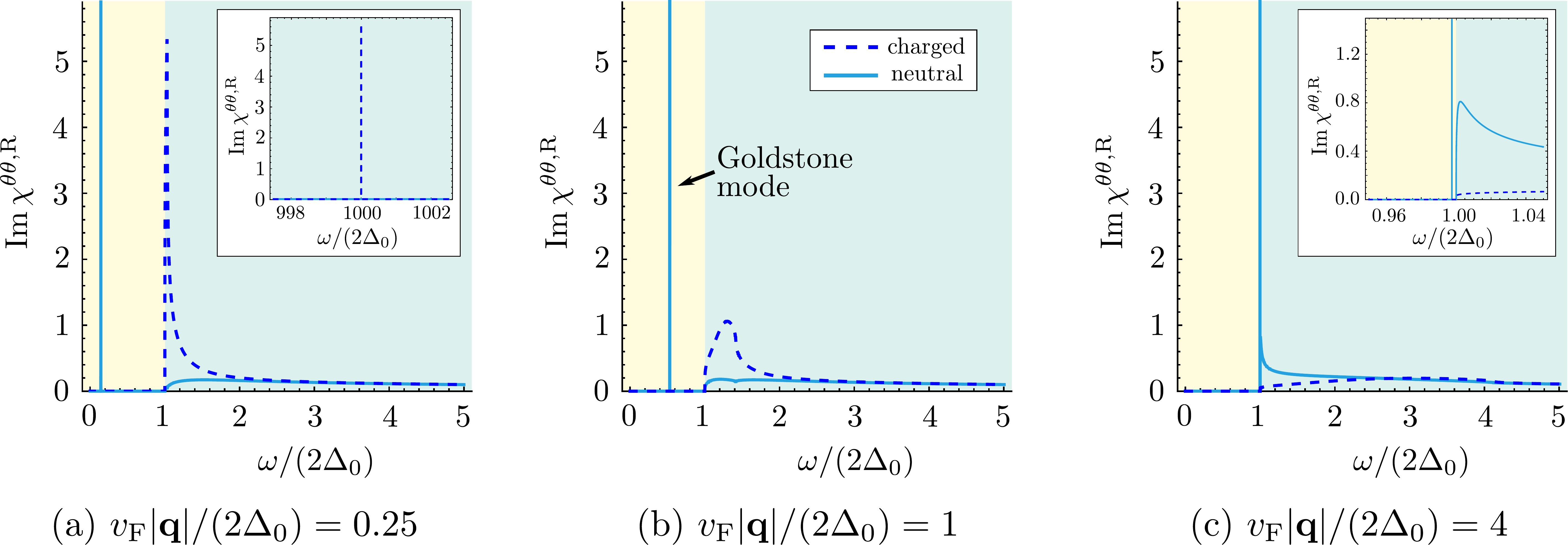}\caption{Phase fluctuation spectrum of a neutral superfluid (light blue line) and a charged superconductor (dark blue dashed line) for
different momenta. Similar to Fig.~\ref{fig:density_plots}, one observes a dramatic reorganization
of the phase excitations in the long-wavelength regime, including 
the Higgs mechanism and the change of the continuum above $2\Delta_0$. 
In the neutral superfluid, the Goldstone mode is shifted closer to $\omega=2\Delta_0$ with increasing momentum; however, it does not merge with the emerging peak of the quasiparticle continuum close to $2\Delta_0$ [see the inset in~(c)]. In the charged superconductor, the Goldstone mode disappears and spectral weight is shifted to the plasma frequency (which we set to $\omega_\mathrm{pl}=2000\Delta_0$ here). This is shown in the inset of~(a).} 
\label{fig:phase-susy-charged-neutral}
\end{figure*}

In this paper, we focus on fermionic systems, which can be either charged or electrically neutral. We refer to the latter as \emph{neutral superfluids}, while the former are addressed as \emph{charged superconductors}, or simply as superconductors, in the remainder. 
The spectrum of phase and amplitude fluctuations for both cases is shown in Figs.~\ref{fig:density_plots} and~\ref{fig:phase-susy-charged-neutral}: While the amplitude mode is not affected by the presence of the Coulomb interaction, the reorganization of the phase excitations becomes apparent here.

Collective excitations and order-parameter fluctuations are well known
to affect the stability of broken-symmetry solutions. Clearly, at
and below the lower critical dimension, long-wavelength fluctuations
destroy off-diagonal long-range order. At $T=0$, this is the case
for one-dimensional systems, where algebraic order is expected in
neutral superfluids and is weakly modified by the long-range Coulomb interaction.
In higher dimensions, $d>1$, the role of fluctuation corrections
to the zero-temperature gap equation of neutral superfluids has attracted some
interest~\cite{Georges91,vanDongen91,MartinRodero92,Kos04,Eberlein13,Eberlein14}.
While such corrections can be sizable once the coupling constant $\lambda$
of the pairing interaction ceases to be small, the BCS~theory~\cite{BCS57}
remains robust for sufficiently weak coupling. 
A particularly revealing analysis of fluctuation effects was given by Kos, Millis,
and Larkin~\cite{Kos04}. These authors analyzed Gaussian fluctuation
corrections to the zero-temperature gap equation. Taking into account fluctuations
of the amplitude and of the phase of the superconducting order parameter,
Kos \emph{et al.}~\cite{Kos04} find that the effect of such corrections
is indeed small. However, their analysis further reveals that the contributions due to phase and amplitude fluctuations are
not individually small in the case of instantaneous interactions; they cancel only in the equation of state. The results
of Kos \emph{et al.}~\cite{Kos04} apply to neutral superfluids. As
discussed above, phase fluctuations of charged superconductors are
dramatically altered by the Higgs mechanism, while amplitude fluctuations
are, at best, weakly affected. Thus, it is unclear whether and how
such a cancellation could take place if one takes the long-range Coulomb
interaction into account. However, without the cancellation of phase
and amplitude fluctuations significant corrections to the BCS~mean-field theory would occur. Thus, the more general question emerges
of how phase fluctuations are modified in a charged superconductor
at different length and energy scales. This question received recent
attention in the context of time-resolved terahertz spectroscopy~\cite{Matsunaga12}
that was interpreted in terms of resonant excitations of the Higgs
particle~\cite{Matsunaga14}. On the other hand, the careful analysis of charge
excitations in superconductors revealed that the direct observation
of the Higgs particle is significantly more subtle~\cite{Cea16}.
\begin{figure}
\includegraphics[width=\columnwidth]{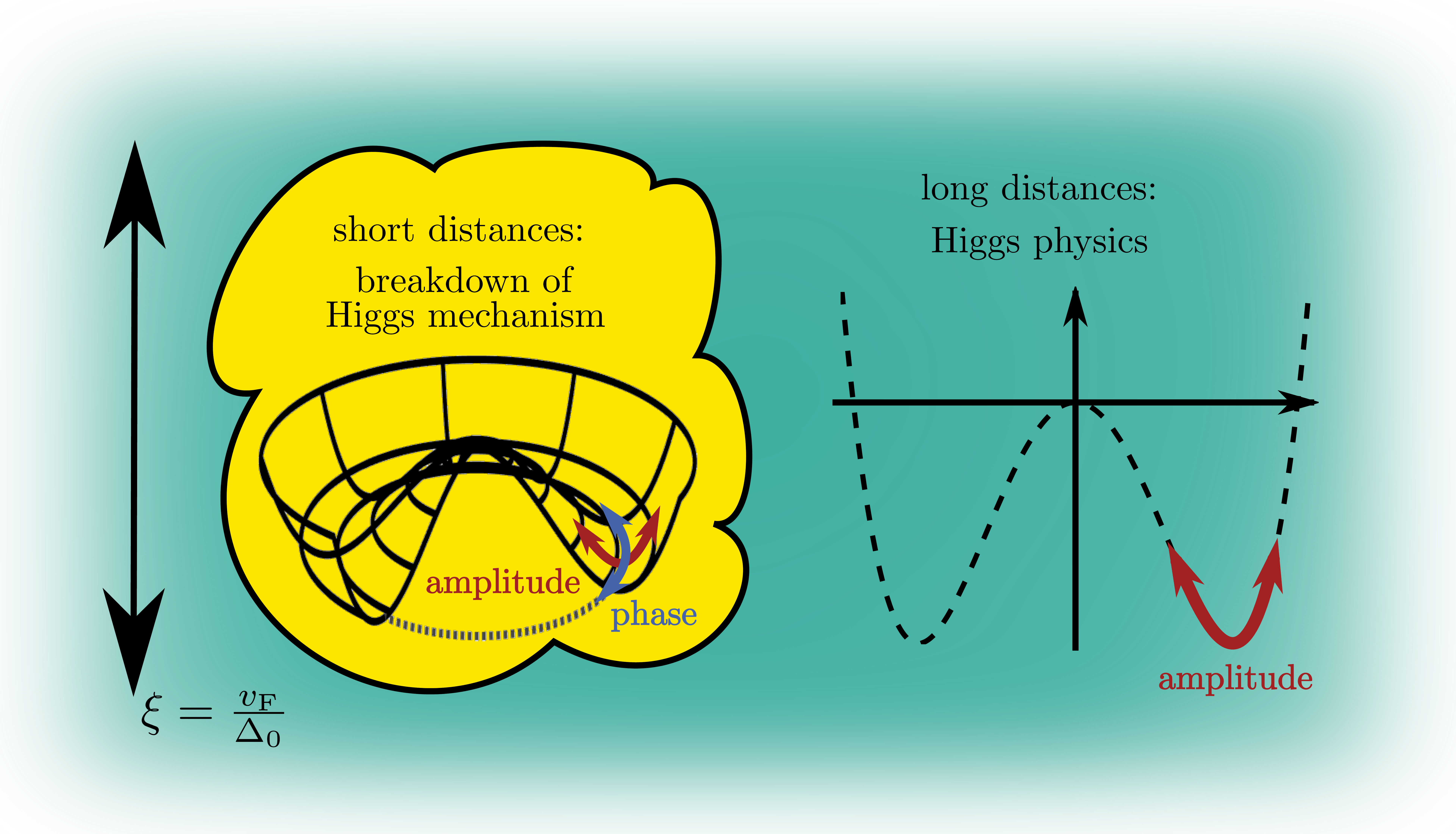}
\caption{While the long-range Coulomb interaction of a superconductor gives
rise to a Higgs mechanism, the composite nature of the Higgs boson,
also known as the Cooper pair, gives rise to a breakdown of the Higgs mechanism
for length scales smaller than the superconducting coherence length $\xi\approx\vf/\Delta_0$, highlighted in yellow here.
The effect takes place only at short distances, and not at high energies,
$\omega>\Delta_0$. It is responsible for the stability of the BCS mean-field
theory. In the long-wavelength regime (indicated in green), the Higgs mechanism is intact; however, quantum fluctuations are negligible for $d>1$.}
\label{fig:blobb}
\end{figure}

In this paper, we investigate the role of Gaussian fluctuations for
the zero-temperature gap equation of a charged superconductor of arbitrary dimensionality~$d$. In the long-wavelength limit, we find the well-known results for the collective modes~\cite{PWA581,PWA582,VolkovKogan-JETP1974,Kulik81},
including the suppression of off-diagonal long-range order at and
below the lower critical dimension~\cite{Hohenberg67}. The strong
coupling between phase fluctuations and plasma excitations eliminates
the long-wavelength Goldstone modes and transfers spectral weight
to the plasma frequency. On the other hand, in the opposite regime
with length scales smaller than the superconducting coherence length
\begin{equation}
\xi=\frac{\vf}{\Delta_{0}}\mathcomma
\end{equation}
excitations of quasiparticles lead to a breakdown of the Higgs mechanism,
where $\vf$ is the Fermi velocity. This is illustrated in Fig.~\ref{fig:blobb}.
We find that the composite nature of the Higgs field implies that
the Higgs mechanism becomes a length-scale-dependent, but not frequency-scale-dependent, phenomenon. As a result, phase and plasmon modes decouple
inside the correlation volume, an effect that is beyond the usual
field-theoretical description of the Higgs mechanism. The weak-coupling
BCS theory of charged superconductors with instantaneous pairing interaction is stable only owing to this
breakdown of the Higgs mechanism for length scales shorter than the
superconducting coherence length $\xi$, where free phase fluctuations
with a large contribution to the gap equation emerge and cancel the
corresponding contribution stemming from amplitude fluctuations. 
Nevertheless, we find that corrections to the gap equation due to short-distance plasma excitations are important unless the dimensionless Coulomb interaction~$\alpha$ [see Eq.~\eqref{eq:dimensionless-Coulomb}] is small compared to~$\lambda^2$, where $\lambda$ is the dimensionless pairing interaction as introduced in~Eq.~\eqref{eq:dimensionless-pairing}. 
\section{Model and Formalism}
We consider a system with the Hamiltonian 
\begin{equation}
\mathcal{H}=\mathcal{H}_{0}+\mathcal{H}_{\mathrm{SC}}+\mathcal{H}_{\mathrm{C}}.
\end{equation}
Here, 
\begin{equation}
\mathcal{H}_{0}=\sum_{\sigma}\int \dd^{d}x\,\psi_{\sigma}^{\dagger}(\mathbf{x})\epsilon(-i\nabla)\psi_{\sigma}(\mathbf{x})
\end{equation}
is the kinetic energy with dispersion $\epsilon(\mathbf{p})$. The local pairing interaction 
\begin{equation}
\mathcal{H}_{\rm{SC}}=-g\int \dd^{d}x\,\psi_{\uparrow}^{\dagger}(\mathbf{x})\psi_{\downarrow}^{\dagger}(\mathbf{x})\psi_{\downarrow}(\mathbf{x})\psi_{\uparrow}(\mathbf{x}) 
\label{eq:pairing}
\end{equation}
with $g>0$ induces superconductivity 
in the $s$-wave channel. While Eq.~\eqref{eq:pairing} is formally
an instantaneous interaction, below we use a characteristic energy
scale $\omega_{0}$ that determines the energy window of the attractive
interaction. For many systems with electron-phonon interaction $\omega_{0}\ll E_\mathrm{F}$ holds, with $E_\mathrm{F}$ being the Fermi energy. A purely
instantaneous interaction corresponds to $\omega_{0}\approx E_\mathrm{F}$.
We will allow for arbitrary values of $\omega_{0}/E_\mathrm{F}$, that is, the
smallness of $\omega_{0}/E_\mathrm{F}$ is not required in the weak-coupling
limit that we consider here. 

Finally, 
\begin{equation}
\mathcal{H}_{\mathrm{C}}=\frac{1}{2}\int \dd^{d}x\int \dd^{d}x^\prime\,\delta n(\mathbf{x})V(\mathbf{x}-\mathbf{x}^\prime)\delta n(\mathbf{x}^\prime)
\end{equation}
describes the electron-electron Coulomb interaction. Here, $\delta n(\mathbf{x})=\sum_{\sigma}\psi_{\sigma}^{\dagger}(\mathbf{x})\psi_{\sigma}(\mathbf{x})-n_{0}$
denotes the electron charge measured relative to the positive ionic background charge~$n_0$.
The unscreened Coulomb interaction $V(\mathbf{x})=e^{2}/|\mathbf{x}|$
will be used for all dimensions~$d$; that is, electric field lines are considered
to extend in three spatial dimensions even for a one- or two-dimensional
superconductor. For $d>1$, the Fourier transform of the Coulomb interaction is given by $V(\mathbf{q})=(4\pi)^{\frac{d-1}{2}}e^{2}\Gamma(\frac{d-1}{2})/|\mathbf{q}|^{d-1}$. For $d=1$, it holds that $V(\mathbf{q})=2e^{2}K_{0}(a|\mathbf{q}|)$,
where~$a$ is the lateral dimension of the wire. For our subsequent analysis,
it is useful to rewrite the Coulomb interaction as $2\dosf V(\mathbf{q})=\frac{\alpha}{\delta}(\frac{1}{2}\xi|\mathbf{q}|)^{1-d}$, with
\begin{equation}
 \delta=\left(\frac{\Delta_{0}}{E_\mathrm{F}}\right)^{d-1}\mathcomma
\end{equation}
$\dosf$ being the density of states at the Fermi energy, 
the dimensionless strength of the Coulomb interaction 
\begin{equation}
\alpha=c_{d}\frac{e^{2}\kf}{E_\mathrm{F}} \mathcomma \label{eq:dimensionless-Coulomb}
\end{equation}
and the numerical coefficient $c_{d}=\frac{2^{2-d}\Gamma\left(\frac{d-1}{2}\right)}{\sqrt{\pi}\Gamma\left(\frac{d}{2}\right)}$.
It holds that $c_{2}=1$ and $c_{3}=\frac{1}{\pi}$. 
Notice that phenomena like the corrections of the pairing interaction due to short-distance Coulomb interactions are assumed to be included in the interaction~$g$. Thus, the pseudopotential~$\mu^\ast$ of Ref.~\onlinecite{Morel62} is already absorbed in the definition of the interaction~$\lambda$ of Eq.~\eqref{eq:dimensionless-pairing}. 

In what follows, we consider the partition function 
\begin{equation}
\mathcal{Z}=\int \mathcal{D}[\bar{\psi},\psi]\,e^{-\mathcal{S}[\bar{\psi},\psi]}\mathcomma
\end{equation}
with the action 
\begin{equation}
\mathcal{S}[\bar{\psi},\psi]=\int_{0}^{\beta}\dd\tau\,\left(\sum_{\sigma}\int \dd^{d}x\,\bar{\psi}_{\sigma}(\mathbf{x})\partial_{\tau}\psi_{\sigma}(\mathbf{x})+\mathcal{H}\right)
\end{equation}
and appropriate source fields to determine observables. We decouple
$\mathcal{H}_{\mathrm{SC}}$ via a Hubbard-Stratonovich transformation in
the Cooper channel and, likewise, $\mathcal{H}_{\mathrm{C}}$ in the plasmon channel.
This introduces a superconducting pairing field $\Delta\left(\mathbf{x},\tau\right)$
and a plasmon field $\phi(\mathbf{x},\tau)$, respectively.
At the saddle point, we obtain the BCS gap equation that determines the mean-field gap~$\Delta_{0}$ as well as the Hartree-Fock shift to the single-particle states
due to the Coulomb interaction. Collective modes can be analyzed by
allowing for fluctuations, 
\begin{equation}
\Delta(\mathbf{x},\tau)=[\Delta_{0}+\eta(\mathbf{x},\tau)]e^{i\theta(\mathbf{x},\tau)}\label{eq:radial-gauge}\mathperiod
\end{equation}
We consider Gaussian fluctuations in $\eta$, $\theta$, and the plasmon
field~$\phi$. 
We performed the analysis in both the Cartesian gauge $\Delta(\mathbf{x},\tau)=\Delta_0+\eta(\mathbf{x},\tau)+i\Delta_0\theta(\mathbf{x},\tau)$ and the radial gauge of Eq.~\eqref{eq:radial-gauge}
and obtained identical results. In what follows we use the Cartesian
gauge. 
The three degrees of freedom are then conveniently combined
into 
\begin{equation}
\Phi=(\eta,-\Delta_{0}\theta,i\phi)\mathperiod \label{eq:components}
\end{equation}

To analyze Gaussian fluctuations, we integrate out gapped fermions
and obtain the partition function 
\begin{equation}
\mathcal{Z}=\int \mathcal{D}[\bar{\Phi},\Phi]\, e^{-\mathcal{S}_\mathrm{eff}[\bar{\Phi},\Phi]}\mathcomma 
\end{equation}
governed by the effective action in terms of the collective bosonic degrees of freedom:  
\begin{equation}
\mathcal{S}_\mathrm{eff}[\bar{\Phi},\Phi] = \int_{q}\left(\frac{\left|\phi_{q}\right|^{2}}{2V(\mathbf{q})}+\frac{\left|\Delta_{q}\right|^{2}}{g}\right)-in_{0}\phi_{0}-{\rm tr}\ln\left(-{\cal G}^{-1}\right).
\end{equation}
Here, $q=\left(\mathbf{q},\omega_{n}\right)$ combines momenta and
bosonic Matsubara frequencies $\omega_{n}=2n\pi T$ and we use the short-hand notation $\int_{q}\ldots= T\sum_{n}\int\dd^{d}q/(2\pi)^{d}\ldots$.
As before, $n_0$ is the homogeneous charge density. 
In addition, $\mathcal{G}_{k,k'}$ is the fermionic propagator in Nambu representation, with $k=\left(\mathbf{k},\nu_{n}\right)$ and fermionic Matsubara
frequencies $\nu_{n}=\left(2n+1\right)\pi T$. ${\cal G}_{k,k'}$
depends on the collective fields contained in the vector $\Phi_{q}$ via
\begin{equation}
\mathcal{G}_{k,k'}^{-1}=\Big[\mathcal{G}^{(0)}_{k}\Big]^{-1}\delta_{k,k'}-\Phi_{k-k'}\cdot\boldsymbol{\tau},
\end{equation}
where $\boldsymbol{\tau}$ is the vector of Pauli matrices in Nambu
space. In addition, 
\begin{equation}
\left[\mathcal{G}_{k}^{(0)}\right]^{-1}\hspace{-0.7em}=i\nu_{n}\tau_{0}-\epsilon_{\mathbf{k}}\tau_{3}-\Delta_{0}\tau_{1}
\end{equation}
is the Green's function of the ordered superconductor with gap $\Delta_{0}$. 

Our goal is to evaluate the partition function $\mathcal{Z}$ at zero temperature
and determine the ground-state energy $E(\Delta_{0})$.
This enables us to analyze the equation of state 
\begin{equation}
\frac{\dd E}{\dd\Delta_{0}^{2}}=0
\end{equation}
that determines the superconducting gap $\Delta_{0}$. 
\section{Gaussian Fluctuations in Charged Superconductors}
Let 
\begin{equation}
\lambda=g\dosf\label{eq:dimensionless-pairing}
\end{equation}
be the dimensionless pairing interaction of the problem. 
At the level of Gaussian fluctuations, we obtain 
\begin{equation}
E=E_{\mathrm{BCS}}+\frac{V}{2}\int_{q}\ln\det\mathcal{D}_{q}^{-1} 
\label{eq:gsenergy}
\end{equation}
for the ground-state energy. In the limit~$\lambda\ll1$, the first term is the well-known ground-state energy of a superconductor
within BCS theory, 
\begin{equation}
E_{\mathrm{BCS}}=E_{0}+V\dosf\Delta_{0}^{2}\left(\ln\frac{\Delta_{0}}{2\omega_{0}}-\frac{1}{2}+\frac{1}{\lambda}\right)\mathcomma 
\end{equation}
where $E_{0}$ is the energy at $\Delta_{0}=0$, $\omega_{0}$ denotes the cutoff
energy of the pairing interaction, and $V$~refers to the volume. Minimizing $E_{\mathrm{BCS}}$
with respect to $\Delta_{0}$ yields the BCS result $\Delta_{0,\mathrm{BCS}}=2\omega_{0}e^{-1/\lambda}$. 

The second term in Eq.~\eqref{eq:gsenergy} is the fluctuation correction.
It is expressed in terms of the $3\times3$ matrix of the fluctuation
propagator $(\mathcal{D}_{q})_{ij}=-\langle T\Phi_{i,q}\Phi_{j,-q}\rangle $
with the inverse 
\begin{equation}
\mathcal{D}_{q}^{-1}=\left(\begin{matrix}
\frac{1}{g} & 0 & 0\\
0 & \frac{1}{g} & 0\\
0 & 0 & \frac{1}{2V(\mathbf{q})}
\end{matrix}\right)-\Pi_{q}\mathperiod
\end{equation}
The components refer to fluctuations of the amplitude, phase, and plasmon field, as well as their mixing. The matrix of the polarization function is 
\begin{equation}
(\Pi_{q})_{ij}=\int_{k}\tr\left(\mathcal{G}_{k}^{(0)}\tau_{i}\mathcal{G}_{k+q}^{(0)}\tau_{j}\right)\mathperiod\label{eq:polarization}
\end{equation}
For the gap equation $\frac{\dd E}{\dd\Delta_{0}^{2}}=0$ it follows that 
\begin{equation}
\frac{1}{V\dosf}\frac{\dd E}{\dd\Delta_{0}^{2}}=\frac{1}{\lambda}-\ln\frac{2\omega_{0}}{\Delta_{0}}+\int_{q}\chi_{q}\mathcomma 
\label{eos}
\end{equation}
where we introduced 
\begin{equation}
\chi_{q}=\frac{\frac{\dd}{\dd\Delta_{0}^{2}}\det\mathcal{D}_{q}^{-1}}{\dosf\det{\cal D}_{q}^{-1}}\mathperiod
\label{eq:fluctuation propagator}
\end{equation}
As long as the last term in Eq.~\eqref{eos} is of order unity or smaller,
the BCS~theory is justified for small~$\lambda$. The authors
of Ref.~\onlinecite{Kos04} find for a neutral superfluid that phase
and amplitude fluctuations each give rise to corrections $\propto\ln\frac{\omega_{0}}{\Delta_{0}}$.
Thus, individually, these corrections cannot be neglected. The stability
of the BCS theory is ensured by the fact that these two large contributions cancel each other exactly. 
This analysis can be generalized to pairing symmetries other than $s$~wave~\cite{ParamekantiEtAl-PRB2000,BarlasVarma-PRB2013,HoyerPhD}: Even though several amplitude and phase modes can arise as a result of the more complex structure of the order parameter, we checked that they always come in equal numbers such that, ultimately, the polarization matrix takes the form of Eq.~\eqref{eq:polarization}. 
Hence, owing to the same cancellation effect, the robustness of the corresponding mean-field theory is ensured. 

In order to proceed, we need to evaluate the matrix elements $(\Pi_{q})_{ij}$
of the polarization function of Eq.~\eqref{eq:polarization} as they determine the bosonic propagators~$(\mathcal{D}_{q})_{ij}$ and the equation
of state. If one is interested in only the long-wavelength behavior
of collective modes, it is sufficient to analyze $(\Pi_{q})_{ij}$ for
wavelengths beyond the coherence length and frequencies below the
superconducting gap, i.\,e., for $|\mathbf{q}|\ll2\xi^{-1}$
and $|\omega|\lesssim2\Delta_{0}$. However, for the analysis
of the equation of state in Eq.~\eqref{eos}, we need to analyze the
particle-particle and particle-hole excitation spectrum for all relevant
momenta~$0<\vf\left|\mathbf{q}\right|<\omega_{0}$ and frequencies~$0<\omega<\omega_{0}$. To this end, we closely follow the approach
and notation of Refs.~\onlinecite{Kos04,Vaks62}. 

We first determine the polarization function $(\Pi_{q})_{ij}$ of Eq.~\eqref{eq:polarization}.
The integration over fermionic momenta and frequencies~at $T=0$ is performed using 
\begin{equation}
\int_{k}\ldots = \dosf\int \dd\epsilon\int\frac{\dd\nu}{2\pi}\int_{\Omega}\ldots\mathcomma 
\end{equation}
where $\epsilon$ is the fermionic energy and $\nu$ is the fermionic Matsubara frequency. In addition, $\int_{\Omega}\ldots=\frac{1}{\Omega_{d}}\int \dd\Omega\ldots$
refers to the integration over the direction of the fermionic momentum
with $\vf\mathbf{k}=\epsilon\mathbf{\boldsymbol{\Omega}}$ and unit
vector $\boldsymbol{\Omega}$. $\Omega_{d}$ is the volume of a $d$-dimensional
unit sphere such that $\int_{\Omega}1=1$. We assume a constant density
of states; that is, the momentum integration is confined to $|\mathbf{k}|<\Lambda\ll k_{\mathrm{F}}$.
Let $\mathbf{q}$ be the external bosonic momentum while $\mathbf{k}$
is the running fermionic momentum in the evaluation of the $(\Pi_{q})_{ij}$.
For an isotropic spectrum, the integrand depends on the momentum direction
through $\cos\angletheta=\frac{\mathbf{k}\cdot\mathbf{q}}{|\mathbf{k}||\mathbf{q}|}$.
For the angular integration, we used that $\int_{\Omega}f(\cos\angletheta)=\tilde{c}_{d}\int_{0}^{1}\dd\mu\,(1-\mu^{2})^{\frac{d-3}{2}}f(\mu)$ holds in $d$~dimensions, with prefactor 
$\tilde{c}_{d}=2\operatorname{\Gamma}\left(\frac{d}{2}\right)/\left(\sqrt{\pi}\operatorname{\Gamma}\left(\frac{d-1}{2}\right)\right)$. 

Following Vaks~\emph{et al.}~\cite{Vaks62}, one can show that the external
momentum and frequency dependence frequently enters through the combination
\begin{equation}
r=\sqrt{v^{2}+u^{2}\cos^{2}\angletheta} \mathcomma 
\label{eq:rofvu}
\end{equation}
with $v=\omega/(2\Delta_{0})$ and $u=\frac{1}{2}\xi|\mathbf{q}|$.
Performing the integration over $\epsilon$ and $\nu$ using the Feynman
parametrization, we then obtain the following results for the polarization
function with a remaining integration over the angle:
\begin{align}
\Pi_{11} & =  \dosf\left(\ln\frac{2\omega_{0}}{\Delta_{0}}-\int_{\Omega}\frac{\arsinh(r)\sqrt{r^{2}+1}}{r}\right) \mathcomma  \\
\Pi_{22} & =  \dosf\left(\ln\frac{2\omega_{0}}{\Delta_{0}}-\int_{\Omega}\frac{\arsinh(r)r}{\sqrt{r^{2}+1}}\right) \mathcomma \\
\Pi_{33} & =-  \dosf\int_{\Omega}\left(\frac{\arsinh(r)v^{2}}{r^{3}\sqrt{r^{2}+1}}+\frac{u^{2}\cos^2\varphi}{r^{2}}\right)\mathperiod
\end{align}
The first two expressions are the same as in Ref.~\onlinecite{Kos04} for
neutral superfluids. Notice that the frequency and momentum dependence of $\Pi_{33}$ (which corresponds
to plasmon excitations) cannot be expressed solely in terms of the combination $r$ introduced in Eq.~\eqref{eq:rofvu}.

If we consider a particle-hole-symmetric superconductor, i.\,e., a system
with a constant density of states $\rho(\epsilon)\approx\dosf$,
the amplitude mode decouples from the other two degrees of freedom,
i.\,e., $\Pi_{1j}=\Pi_{j1}=0$ for $j\in\{2,3\}$. Allowing for
a violation of particle-hole symmetry gives rise to only small
corrections of the order of $\frac{\dosf^\prime}{\dosf}\Delta_{0}$. Therefore, we assume particle-hole symmetry in the remainder of the paper, and as a consequence, all off-diagonal matrix elements
vanish except for 
\begin{equation}
\Pi_{23}=-\dosf\int_{\Omega}\frac{{\rm arcsinh}\left(r\right)v}{r\sqrt{r^{2}+1}}.
\end{equation}
This phase-plasmon coupling changes the dynamics of the phase fluctuations
and charge excitations in the superconductor due to the Coulomb interaction
and is responsible for the Higgs mechanism.

For the collective boson propagator~$\chi_{q}$ introduced in Eq.~\eqref{eq:fluctuation propagator} it follows that  
\begin{equation}
\chi_{q}=\chi_{{\rm amp},q}+\chi_{{\rm pp},q}\mathcomma 
\end{equation}
that is, it can be split into a contribution due to amplitude fluctuations, 
\begin{equation}
\chi_{\mathrm{amp}}=\frac{1}{\dosf\mathcal{D}_{11}^{-1}}\frac{\dd\mathcal{D}_{11}^{-1}}{\dd\Delta_{0}^{2}}\mathcomma 
\end{equation}
and a contribution associated with the coupled phase-plasmon fluctuations, 
\begin{equation}
\chi_{\mathrm{pp}}=\frac{\frac{\dd\mathcal{D}_{22}^{-1}}{\dd\Delta_{0}^{2}}\mathcal{D}_{33}^{-1}+\frac{\dd\mathcal{D}_{33}^{-1}}{\dd\Delta_{0}^{2}}\mathcal{D}_{22}^{-1}-2\frac{\dd\mathcal{D}_{23}^{-1}}{\dd\Delta_{0}^{2}}\mathcal{D}_{23}^{-1}}{\dosf(\mathcal{D}_{22}^{-1}\mathcal{D}_{33}^{-1}-\mathcal{D}_{23}^{-1}\mathcal{D}_{23}^{-1})}\mathperiod
\label{eq:chi pp}
\end{equation}
Obviously, if we take the limit $\mathcal{D}_{23}^{-1}\rightarrow0$ and  $\mathcal{D}_{33}^{-1}\rightarrow0$, we recover the limit of a neutral superfluid: $\chi_{\mathrm{pp}}\rightarrow\chi_{\mathrm{ph}}=\frac{1}{\mathcal{D}_{22}^{-1}}\frac{\dd\mathcal{D}_{22}^{-1}}{\dd\Delta_{0}^{2}}$.
For the evaluation of the derivatives with respect to $\Delta_{0}^{2}$, 
we use $\frac{\dd}{\dd\Delta_0^{2}}\ln\frac{2\omega_{0}}{\Delta_{0}}=-\frac{1}{2\Delta_0^{2}}$
and $\frac{\dd f(r)}{\dd\Delta_0^{2}}=-\frac{1}{2\Delta_0^{2}}r\frac{\dd f}{\dd r}$, which results in 
\begin{align}
\frac{\dd\mathcal{D}_{11}^{-1}}{\dd\Delta_{0}^{2}} & =  \frac{\dosf}{2\Delta_{0}^{2}}\int_{\Omega}\frac{\arsinh(r)}{r\sqrt{1+r^{2}}} \mathcomma  \\
\frac{\dd\mathcal{D}_{22}^{-1}}{\dd\Delta_{0}^{2}} & =  \frac{\dosf}{2\Delta_{0}^{2}}\int_{\Omega}\left(\frac{1}{1+r^{2}}-\frac{\arsinh(r)r}{(1+r^{2})^{3/2}}\right)\mathcomma  \\
\frac{\dd\mathcal{D}_{33}^{-1}}{\dd\Delta_{0}^{2}} & =  \frac{\dosf}{2\Delta_{0}^{2}}\int_{\Omega}\frac{v^{2}}{r^{2}}\left(\frac{\arsinh(r)(1+2r^{2})}{r(1+r^{2})^{3/2}}-\frac{1}{1+r^{2}}\right) \mathcomma \\
\frac{\dd\mathcal{D}_{23}^{-1}}{\dd\Delta_{0}^{2}} & =  \frac{\dosf}{2\Delta_{0}^{2}}v\int_{\Omega}\left(\frac{\arsinh(r)r}{(1+r^{2})^{3/2}}-\frac{1}{1+r^{2}}\right) \mathperiod
\end{align}
In what follows, we analyze the integrals that enter the equation
of state in the long-wavelength limit $|\mathbf{q}|<2\xi^{-1}$
and in the opposite limit $2\xi^{-1}<\left|\mathbf{q}\right|<\omega_{0}/\vf$.  We focus on contributions that could amount to corrections comparable to the BCS~theory. Contributions to $\int_q\chi_q$ that are merely of order unity or smaller will be ignored in the following. 

In performing the remaining momentum and frequency integration, we make use of the relation 
\begin{align}
 &\int\frac{\dd\omega}{2\pi}\int\frac{\dd^d q}{(2\pi)^{d}}\,\frac{1}{\dosf\Delta_0^2}\,f\left(\frac{\omega}{2\Delta_0},\frac{\vf|\mathbf{q}|}{2\Delta_0}\right) \nonumber \\ 
 &\qquad =\frac{1}{2\pi}\left(\frac{\Delta_0}{E_\mathrm{F}}\right)^{d-1}\int\dd v\int \dd u\,u^{d-1}f(v,u)\mathcomma 
\end{align}
where we used $E_\mathrm{F}=\frac{1}{2}\vf\kf$, $\vf=\kf/m$, and $\dosf=K_d\kf^{d-1}/\vf$, with $K_d=1/[2^{d-1}\pi^{d/2}\Gamma(d/2)]$. Depending on the regime under consideration, the integration limits are chosen appropriately. 
\subsection{Long-wavelength fluctuations}\label{sec:long-wavelength}
We start our analysis of long-wavelength fluctuations $|\mathbf{q}|\ll2\xi^{-1}$ by
considering, in addition, small frequencies $\omega\ll2\Delta_{0}$. 
This corresponds to regime~I, as indicated in Fig.~\ref{fig:density_plots}.(a).
Thus, we can safely assume that $u$, $v$, and $r$ are all small
compared to unity. 

We first consider amplitude fluctuations, i.\,e., $\chi_{\mathrm{amp}}$.
For small $r$, it holds that $\frac{\dd\mathcal{D}_{11}^{-1}}{\dd\Delta_{0}^{2}}\approx\frac{\dosf}{2\Delta_{0}^{2}}$
and $\mathcal{D}_{11}^{-1}\approx\dosf$. Thus, $\chi_{\mathrm{amp}}\approx\frac{1}{2\dosf\Delta_{0}^{2}}$.
The correction to the equation of state due to low-energy amplitude
fluctuations is thus given by 
\begin{align}
\int_{q}^{<}\chi_{\mathrm{amp},q} & = \int_{0}^{2\Delta_{0}}\frac{\dd\omega}{(2\pi)}\int_{|\mathbf{q}|<2\xi^{-1}}\frac{\dd^{d}q}{(2\pi)^{d}}\frac{1}{2\dosf\Delta_0^2}\nonumber \\
&=\frac{1}{4\pi d}\left(\frac{\Delta_0}{E_\mathrm{F}}\right)^{d-1}\ll 1 \mathcomma 
\end{align}
where the last inequality holds for $d>1$. 
The superscript $<$ of the integral symbol implies that we consider only 
contributions below $2\Delta_0$ in energy and $2\xi^{-1}$ in wave number.
The integral $\int_{q}^{<}\chi_{{\rm amp},q}$ has to be compared
to $\ln(2\omega_{0}/\Delta_{0})$ in Eq.~\eqref{eos}. Obviously,
low-energy and long-wavelength amplitude fluctuations never give
rise to relevant corrections. This is also true at the lower critical dimension
$d=1$, where corrections to $\ln(2\omega_{0}/\Delta_{0})$
are still only of order unity. This is to be expected as these excitations
are gapped below $\omega=2\Delta_{0}$. 

Next, we analyze long-wavelength low-energy contributions due to phase-plasmon
excitations. In the case of a neutral superfluid, the propagator $\chi_{\mathrm{pp}}$
of phase fluctuations is given by $\chi_{\mathrm{pp}}=\frac{1}{\dosf(\omega^{2}+\frac{1}{d}\vf^{2}|\mathbf{q}|^{2})}$,
which yields, after an analytic continuation to the real-frequency axis,
the Goldstone mode $\omega_{\mathrm{G}}=\frac{1}{\sqrt{d}}\vf|\mathbf{q}|$. This situation
changes when we include the Coulomb interaction. Expanding for small
$\omega$ and $|\mathbf{q}|$ now yields 
\begin{equation}
\chi_{\mathrm{pp}}=\frac{V(\mathbf{q})}{\omega^{2}+\frac{n_{0}^{2}}{m^{2}}|\mathbf{q}|^{2}V(\mathbf{q})}\mathperiod
\end{equation}
Then, after analytic continuation to the real axis, we find a pole at the
plasma frequency $\omega_{\mathrm{pl}}=\sqrt{\frac{n_{0}}{m}|\mathbf{q}|^{2}V(\mathbf{q})}$.
As usual, $\omega_{{\rm pl}}$ is finite for $d=3$ and vanishes like
$\omega_{\rm{pl}}=\sqrt{\frac{2\pi e^{2}n_{0}}{m}}|\mathbf{q}|^{1/2}$
for $d=2$. In one-dimensional systems, the plasma frequency is almost
linear, $\omega_{\mathrm{pl}}=\sqrt{\frac{2e^{2}n_{0}}{m}\ln\frac{2e^{-\gamma_{E}}}{a|\mathbf{q}|}}|\mathbf{q}|$. 

The correction in the equation of state due to low-energy phase-plasmon
fluctuations is 
\begin{equation}
\int_{q}^{<}\chi_{\mathrm{pp},q}=\int_{0}^{2\Delta_{0}}\frac{\dd\omega}{(2\pi)}\int_{0}^{2\xi^{-1}}\frac{\dd^{d}q}{(2\pi)^{d}}\,\frac{V(\mathbf{q})}{\omega^{2}+\frac{n_{0}^{2}}{m^{2}}|\mathbf{q}|^{2}V(\mathbf{q})}\mathperiod
\end{equation}
For dimensions $d>1$, this integral equals {$\sqrt{\alpha}/{[16(d-1)]}(\Delta_0/E_\mathrm{F})^{(d-1)/2}$}, and hence, corrections to the equation of state are negligible. For $d=1$, the
integral diverges like $(\ln|\mathbf{q}|)^{\frac{3}{2}}$ at the lower limit.
Thus, phase-plasmon fluctuations suppress long-range order for $d=1$.
The same is true for neutral superfluids, where the integral diverges
as $\ln|\mathbf{q}|$.

To summarize, long-wavelength contributions from phase-plasmon and
amplitude fluctuations yield the expected behavior. Significant corrections
occur only at the lower critical dimension, which is $d_{{\rm lc}}=1$
for a superconductor at $T=0$.

\subsection{Short-wavelength fluctuations}\label{sec:short-wavelength}
Next, we analyze the behavior in the opposite limit, that is, in the regime of short-wavelength fluctuations, by which we refer to distances shorter than the correlation length, i.\,e., $2\xi^{-1}\ll|\mathbf{q}|\ll\kf$, hence not including the regime where lattice effects become important. In this regime, it is possible to assume that $r\gg1$ is satisfied and expand all relevant expressions for large $r$. Since this is the case not only when the momenta of collective bosonic excitations are
large compared to $2\Delta_0/\vf$ but also when the frequencies are large compared to $2\Delta_0$, we treat both cases in this section. Our analysis, however, reveals that the former is physically more interesting. 

We first analyze amplitude fluctuations. In this case, it holds that $\mathcal{D}_{11}^{-1}=\dosf\int_{\Omega}\ln(2r)$
and $\frac{\dd\mathcal{D}_{11}^{-1}}{\dd\Delta_{0}^{2}}=\frac{\dosf}{2\Delta_{0}^{2}}\int_{\Omega}\frac{\ln(2r)}{r^{2}}$.
Hence the collective propagator of amplitude modes is given by 
\begin{equation}
\int_{q}^{>}\chi_{\mathrm{amp},q}=\int\frac{\dd\omega}{2\pi}\int\frac{\dd^{d}q}{(2\pi)^{d}}\,\frac{1}{2\dosf\Delta_{0}^{2}}\frac{\int_{\Omega}\frac{\ln\left(2r\right)}{r^{2}}}{\int_{\Omega}\ln\left(2r\right)},\label{eq:chiamp high}
\end{equation}
where we do not explicitly show the dependence of $r$ on $\omega$
and $|\mathbf{q}|$ as given by Eq.~\eqref{eq:rofvu}. The superscript $>$ of the integral
symbol implies that we consider only contributions that are either above
$2\Delta_0$ in energy or $2\xi^{-1}$ in wave number. 
Of course, for very large bosonic frequencies, we have to cut off the
integral at the scale $\omega_{0}$. 
It is now useful
to consider two different regimes $\omega\gtrless \vf |\mathbf{q}|$ for the
integrand, i.\,e., $v\gtrless u$. For $v>u$ [see regime~II in Fig.~\ref{fig:density_plots}.(a)] the angular
integration is trivial, that is, $\int_{\Omega}f(r)\approx f(v)$.
Performing the momentum and frequency integration with the limits
$\int_{1}^{\frac{\omega_{0}}{2\Delta_{0}}}\dd v\int_{0}^{v}\dd u\, u^{d-1}\ldots$, we obtain the result $\chi_{\mathrm{amp}}^{(\omega>\vf|\mathbf{q}|)}\approx(\omega_{0}/E_{\mathrm{F}})^{d-1}$ for $d>1$. 
In the opposite limit of $u>v$ [see regime~III in Fig.~\ref{fig:density_plots}.(a)] holds that $\int_{\Omega}\frac{\ln(2r)}{r^{2}}\approx\frac{\ln v}{uv}$
and $\int_{\Omega}\ln(2r)=\ln (2u)$. If we
now perform the integration with the opposite order of limits, we
find 
\begin{equation}
\int_q^>\chi_\mathrm{amp}^{(\omega<\vf|\mathbf{q}|)}=\frac{1}{8\pi(d-1)}\left(\frac{\omega_0}{E_\mathrm{F}}\right)^{d-1}\ln\left(\frac{\omega_0}{\Delta_0}\right)\mathperiod 
\label{eq:amplitude-short-mom}
\end{equation}
Thus, the impact of amplitude fluctuations is dominated by regime~III, defined by momenta for which $|\mathbf{q}|\gg2\xi^{-1}$ and $|\mathbf{q}|>\omega/\vf$ hold. 
Note that the result~\eqref{eq:amplitude-short-mom} originates from the regime $1<v<u$.  
However, we checked that, similar to the long-wavelength limit, the regime $v<1<u$ yields only small corrections of order $(\omega_0/E_\mathrm{F})^{d-1}$. 
The qualitatively distinct behavior of the fluctuation spectrum with respect to energy and momentum can roughly be captured by
\begin{equation}
 \chi_{\mathrm{amp},q}\sim \frac{1}{\omega \max(\omega,\vf|\mathbf{q}|)}\mathcomma 
\end{equation}
which explains why the equation of state is dominated by $\vf|\mathbf{q}|>\omega$. 
This is qualitatively different from Lorentz-invariant theories of Higgs fields. 

For an instantaneous, i.\,e., electronic
pairing mechanism with $\omega_{0}\sim E_{\mathrm{F}}$, the correction
due to amplitude fluctuations is 
\begin{equation}
\int_{q}^{>}\chi_{\mathrm{amp},q}^{(\omega<\vf|\mathbf{q}|)}=\frac{1}{8\pi(d-1)}\ln\left(\frac{\omega_{0}}{\Delta_{0}}\right)\mathperiod
\label{eq:Log}
\end{equation} 
If this were the only correction to the mean-field theory, clearly, the BCS approach would not be controlled for small $\lambda$. In
the case of a neutral superfluid, it is easy to see that phase
fluctuations with large momentum and frequency, i.\,e., for large
$r$, behave as
\begin{equation}
\int_{q}^{>}\chi_{\mathrm{ph},q}=-\int\frac{\dd\omega}{2\pi}\int\frac{\dd^{d}q}{(2\pi)^{d}}\,\frac{1}{2\dosf\Delta_{0}^{2}}\frac{\int_{\Omega}\frac{\ln(2r)}{r^{2}}}{\int_{\Omega}\ln(2r)}\mathperiod
\end{equation}
Comparing this expression with Eq.~\eqref{eq:chiamp high}, we see that
phase fluctuations of neutral superfluids exactly compensate
the $\ln\left(\omega_{0}/\Delta_{0}\right)$ corrections of amplitude
fluctuations. This was the key result of Ref.~\onlinecite{Kos04}. While
amplitude fluctuations tend to suppress the pairing amplitude, phase
fluctuations enhance it. This is consistent with the fluctuation analysis of Ref.~\onlinecite{Eberlein13}. 

Next, we analyze the behavior of charged superconductors. To this end,
we determine the large-$r$ behavior of the coupled phase-plasmon
propagator $\chi_{\mathrm{pp}}$ of Eq.~\eqref{eq:chi pp}. In this limit 
\begin{align}
\mathcal{D}_{22}^{-1} & = \dosf\int_{\Omega}\ln(2r) \mathcomma \\
\mathcal{D}_{33}^{-1} & =\dosf  \left(\frac{\delta}{\alpha}u^{d-1}+\int_{\Omega}\left[1-\frac{v^{2}}{r^2}\left(1-\frac{\ln\left(2r\right)}{r^{2}}\right)\right]\right) \mathcomma \\
\mathcal{D}_{23}^{-1} & = \dosf v\int_{\Omega}\frac{\ln(2r)}{r^{2}}\mathcomma 
\end{align}
as well as 
\begin{align}
\frac{\dd\mathcal{D}_{22}^{-1}}{\dd\Delta_{0}^{2}} & = -\frac{\dosf}{2\Delta_{0}^{2}}\int_{\Omega}\frac{\ln(2r)}{r^{2}} \mathcomma  \\
\frac{\dd\mathcal{D}_{33}^{-1}}{\dd\Delta_{0}^{2}} & = \frac{\dosf}{2\Delta_0^2}\int_{\Omega}\frac{2 v^2\ln(2r)}{r^4}\mathcomma  \\
\frac{\dd\mathcal{D}_{23}^{-1}}{d\Delta_{0}^{2}} & = \frac{\dosf}{2\Delta_{0}^{2}}v\int_{\Omega}\frac{\ln(2r)}{r^{2}}\mathperiod
\end{align}

We first analyze regime~II, where the frequency $\omega$ dominates
over the momentum $\vf|\mathbf{q}|$, i.\,e., where $v\gg u$. In this case, 
\begin{align}
\mathcal{D}_{22}^{-1} & = \dosf\ln(2v)\mathcomma  \\
\mathcal{D}_{33}^{-1} & = \dosf\left(\frac{\delta}{\alpha}u^{d-1}+\frac{\ln(2v)}{v^{2}}\right) \mathcomma\\
\mathcal{D}_{23}^{-1} & = \dosf\frac{\ln(2v)}{v} \mathcomma 
\end{align}
as well as
\begin{align}
\frac{\dd\mathcal{D}_{22}^{-1}}{\dd\Delta_{0}^{2}} & = -\frac{\dosf}{2\Delta_{0}^{2}}\frac{\ln(2v)}{v^{2}} \mathcomma \\
\frac{\dd\mathcal{D}_{33}^{-1}}{\dd\Delta_{0}^{2}} & = \frac{\dosf}{2\Delta_{0}^{2}}\frac{2\ln(2v)}{v^{2}} \mathcomma \\
\frac{\dd\mathcal{D}_{23}^{-1}}{\dd\Delta_{0}^{2}} & = \frac{\dosf}{2\Delta_{0}^{2}}\frac{\ln(2v)}{v} \mathperiod
\end{align}
This yields 
\begin{equation}
\chi_{\mathrm{pp}}^{(\omega>\vf|\mathbf{q}|)}=-\frac{1}{2\dosf\Delta_0^2}\frac{1}{v^2}
\label{chi pp v}
\end{equation}
for the phase-plasmon propagator. 
Performing the integration over momenta and frequencies assuming that 
$\omega>\vf|\mathbf{q}|$ finally gives 
\begin{equation}
\int_{q}^{>}\chi_{\mathrm{pp},q}^{(\omega>\vf|\mathbf{q}|)}=-\frac{1}{4\pi d}\left(\frac{\omega_0}{E_\mathrm{F}}\right)^{d-1}\mathperiod
\end{equation}
In this limit of frequencies larger than momenta, there is no relevant correction
to the equation of state due to collective phase-plasmon excitations. It is also noteworthy that the structure of $\chi_{\mathrm{pp}}$ as given in Eq.~\eqref{chi pp v}
heavily relies on the strong coupling between phase and plasmon degrees
of freedom. The Higgs mechanism is fully intact in this high-frequency
but long-distance regime. 

\begin{figure*}[t]
\includegraphics[width=\textwidth]{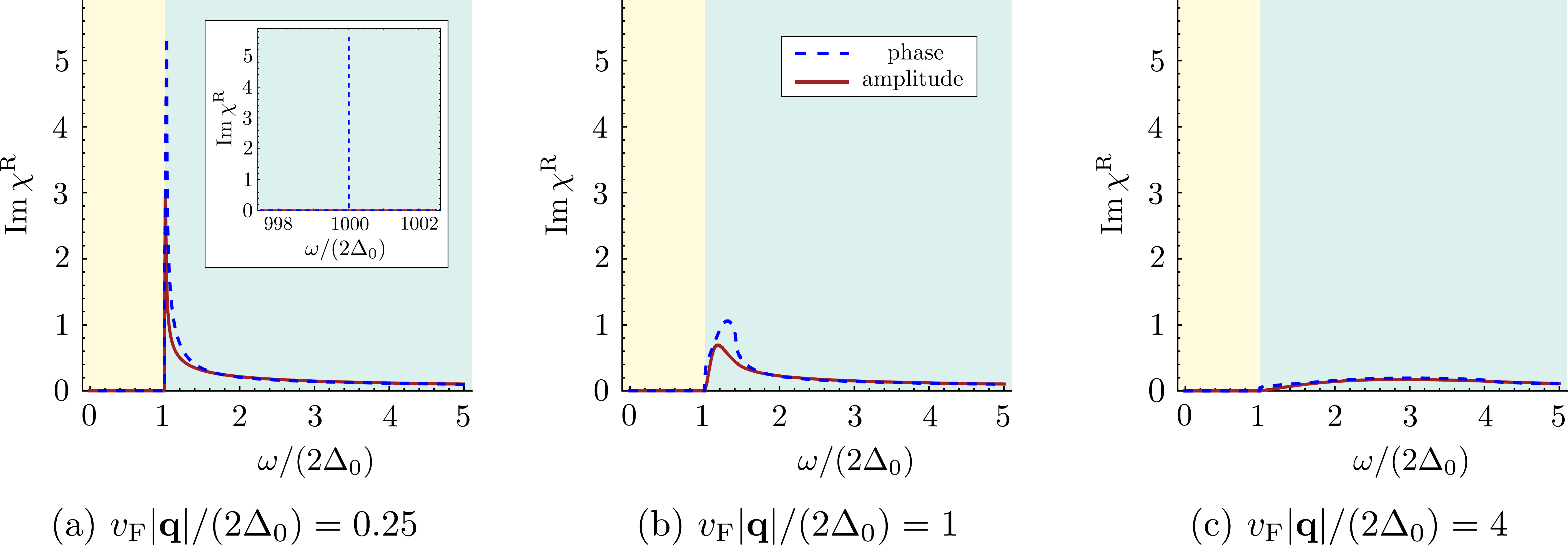}
\caption{Spectrum of phase (blue dashed line) and amplitude (solid red line) fluctuations in a charged superconductor.
While both excitations are very different in character for neutral
superfluids [compare Figs.~\ref{fig:density_plots} and~\ref{fig:phase-susy-charged-neutral}], they behave increasingly similar in charged systems at
large momenta as illustrated here, the main difference being the spectral weight at the plasma frequency shown in the inset of~(a).}
\label{fig:phase-amplitude-susy-charged}
\end{figure*}
We finally focus on the case of $\vf|\mathbf{q}|>\omega$, where amplitude
fluctuations yield significant corrections to the equation of state.
Thus, we analyze the case where $u\gg v$, i.\,e., regime~III. Note that the subsequent discussion again refers to $1<v<u$. Corrections stemming from the regime $v<1<u$ can be calculated analogously but turn out to be subleading in both neutral superfluids and charged superconductors. In this limit, 
\begin{align}
\mathcal{D}_{22}^{-1} & =  \dosf\ln(2u)\mathcomma  \\
\mathcal{D}_{23}^{-1} & = \dosf\tilde{c}_d\frac{\ln v}{u}\mathcomma \\
\mathcal{D}_{33}^{-1} & =\dosf \left(\frac{\delta}{\alpha}u^{d-1}+1\right) \mathcomma 
\end{align}
as well as 
\begin{align}
\frac{\dd\mathcal{D}_{22}^{-1}}{\dd\Delta_{0}^{2}} & =-\frac{\dosf\tilde{c}_d}{2\Delta_{0}^{2}}\frac{\ln v}{uv} \mathcomma \\
\frac{\dd\mathcal{D}_{23}^{-1}}{\dd\Delta_{0}^{2}} & = \frac{\dosf\tilde{c}_d}{2\Delta_{0}^{2}}\frac{\ln v}{u} \mathcomma  \\
\frac{\dd\mathcal{D}_{33}^{-1}}{\dd\Delta_{0}^{2}} & = \frac{\dosf\tilde{c}_d}{2\Delta_{0}^{2}}\frac{1}{3}\frac{2\ln(v)}{uv} \mathperiod
\end{align}
As a result, the phase-plasmon propagator takes the form 
\begin{equation}
\chi_{\mathrm{pp}}^{(\vf|\mathbf{q}|>\omega)}=\frac{1}{2\dosf\Delta_{0}^2}\left(-\frac{\ln v}{\ln(2u)uv}+\frac{2\alpha\ln(v)}{3uv(\delta u^{d-1}+\alpha)}\right)\mathperiod 
\label{eq:res}
\end{equation}
Now performing the integration over momenta and frequencies, we find
that the first term in $\chi_{\mathrm{pp}}^{(\vf|\mathbf{q}|>\omega)}$
exactly cancels the contribution due to amplitude fluctuations of
Eq.~\eqref{eq:amplitude-short-mom}. This cancellation can also be seen in Fig.~\ref{fig:phase-amplitude-susy-charged}, where
we compare the spectral function of amplitude and phase modes of a
charged superconductor for different momenta. The cancellation of
fluctuations is a consequence of the fact that the coupling between
phase and plasmon modes vanishes in the regime $\vf|\mathbf{q}|>\omega$ and $|\mathbf{q}|>2\xi^{-1}$.
Thus, inside the correlation volume $\xi^{d}$ of the superconductor the Higgs mechanism is inactive. Notice that this statement is valid
for all frequencies, i.\,e., also in the low-energy limit. On the other
hand, for $|\mathbf{q}|<2\xi^{-1}$, it holds that the phase-plasmon coupling is strong,
including at large energies $\omega>2\Delta_{0}$. Thus, there is
a short-distance breakdown of the Higgs mechanism. In fact, we could
have written 
\begin{equation}
\dosf\chi_{\mathrm{pp}}^{(\vf|\mathbf{q}|>\omega)}\approx\mathcal{D}_{22}\frac{\dd\mathcal{D}_{22}^{-1}}{\dd\Delta_{0}^{2}}+\mathcal{D}_{33}\frac{\dd\mathcal{D}_{33}^{-1}}{\dd\Delta_{0}^{2}}
\label{eq:decoupling-pp-propagator}
\end{equation}
to obtain Eq.~\eqref{eq:res}, making the decoupling of charge and phase
fluctuations manifest. 

Finally, we analyze under what condition the correction due to plasmon 
fluctuations, i.\,e., the last term in Eq.~\eqref{eq:decoupling-pp-propagator}, is relevant. Our preceding analysis yields that no relevant
corrections occur in the regime where $\omega>\vf |\mathbf{q}|$ as long as
$\alpha$ is not larger than unity, an assumption that is needed to justify our random-phase approximation treatment of the Coulomb interaction. In the opposite limit, for instantaneous interactions,
\begin{align}
\int_{q}^{>}\chi_q^{(\vf|\mathbf{q}|>\omega)} & = \int_{q}^{>}\left(\chi_{\mathrm{amp},q}^{(\vf|\mathbf{q}|>\omega)}+\chi_{\mathrm{pp},q}^{(\vf|\mathbf{q}|>\omega)}\right)\nonumber \\
& \hspace{-4em} = \frac{\alpha}{4\pi}\int_1^{\frac{E_\mathrm{F}}{2\Delta_0}}\dd u\,u^{d-1}\int_1^u\dd v\,\frac{2\ln(v)}{3uv(u^{d-1}+\frac{\alpha}{\delta})}\mathperiod
\end{align}
The analysis of this integral is straightforward. The leading correction to the BCS theory for $\alpha<1$ is 
\begin{equation}
\int_q\chi_q=\frac{\alpha\tilde{c}_d}{36\pi}\ln^3\left(\frac{E_\mathrm{F}}{2\Delta_0}\right)\mathperiod
\end{equation}
This is negligible compared to $\lambda^{-1}$ only if the dimensionless
Coulomb interaction $\alpha$ is smaller than the square of the dimensionless pairing
interaction, 
\begin{equation}
\alpha\ll\frac{36\pi}{\tilde{c}_d} \lambda^2 =\left\{ \begin{matrix} 18\pi^2\lambda^2 & \text{for } d=2\mathcomma  \\ 36\pi\lambda^2 & \text{for } d=3\mathperiod\end{matrix}\right. 
\label{eq:condition-alpha}
\end{equation}
Thus, for weak pairing~$\lambda$, the restrictions for the strength of the Coulomb interaction are quite severe and may only be fulfilled due to a large numerical prefactor. 
Note that this effect is due to plasma excitations that are changed by the superconducting state. It is different from the pseudopotential effects discussed by Morel and Anderson~\cite{Morel62} that we already included in the definition of~$\lambda$ (see above).  
\section{Conclusions}
In this paper, we investigated the impact of Gaussian fluctuations
on the equation of state of a superconductor, taking into account
the role of the long-range Coulomb interaction on the gap equation.
This investigation was motivated by two observations. (i) In Ref.~\onlinecite{Kos04},
the BCS~mean-field theory of neutral superfluids was shown to be stable
because of a subtle cancellation of amplitude and phase fluctuations
in the equation of state. (ii) The phase excitation spectrum of a charged
superconductor is qualitatively different from neutral superfluids,
which is most evident from the Higgs mechanism that eliminates the Goldstone
mode of the neutral case. This led to the question of whether the
above cancellation is still present in charged superconductors. Our analysis
demonstrated  that, even
for the charged case, the stability of the BCS theory is guaranteed for instantaneous pairing interactions. This is a consequence of a breakdown of the
Higgs mechanism on length scales shorter than the superconducting
coherence length. In this regime with $|\mathbf{q}|>2\xi^{-1}$, plasma excitations
and phase fluctuations decouple. It is the same regime that contributes
to the dominant fluctuations that threaten the robustness of the mean-field theory.
These subtleties are a direct consequence of the fact that the Higgs
field of superconductors is a composite field where two fermions are
bound to form a Cooper pair. The composite nature of the Higgs particle
is then responsible for the vastly different momentum and frequency
dependence of the dynamics of the Higgs field, including the mentioned
breakdown of the Higgs mechanism. 
Notice that the Higgs mechanism is fully intact for energy scales up to the Fermi/plasma energy as long as we consider long-wavelength fluctuations. 
We also demonstrated that plasmon
excitations do lead to significant corrections to the gap equation
unless the dimensionless strength of the Coulomb interaction is small
compared to the square of the attractive pairing interaction of the superconducting
state~[Eq.~\eqref{eq:condition-alpha}]. 
This finding demonstrates that the change in the density fluctuation spectrum of a superconductor is not limited to the energy regime near the superconducting gap. 
Another interesting implication of our analysis is with regard 
to many-body theories that have structural similarities to the BCS~theory, such as density-wave instabilities with perfect nesting. In
this case, it can be shown that the cancellation of transverse and
longitudinal modes is not always guaranteed, leading to important limitations
of the mean-field theory of density-wave systems~\cite{Hoyer17}. 
The diagrammatic representation of the fluctuation corrections to the gap equation (22) along with the discussion of the importance of the different contributions is presented in Ref.~\onlinecite{Hoyer17} for a closely related problem. 
\section*{Acknowledgments}
We are grateful to A.~V.~Chubukov, W.~Metzner, M.~S.~Scheurer, and C.~M.~Varma for helpful discussions. 

%

\end{document}